\documentstyle[aps,prl,floats,epsf,twocolumn]{revtex}
\begin{document}
\newcommand{\dm}       {\Delta m^2}
\newcommand{\sinq}      {\sin^2 2\theta}
\newcommand{\nuebar}      {\bar\nu_{\rm e}}
\draft
\preprint{HEP/123-qed}
\wideabs{
\title{Final results from the Palo Verde Neutrino Oscillation Experiment}
\author{F.~Boehm$^3$, J.~Busenitz$^1$, B.~Cook$^3$, G.~Gratta$^4$,
        H.~Henrikson$^3$, J.~Kornis$^1$, D.~Lawrence$^2$,
        K.B.~Lee$^3$, K.~McKinney$^1$, L.~Miller$^4$,
        V.~Novikov$^3$, A.~Piepke$^{1,3}$, B.~Ritchie$^2$, D.~Tracy$^4$,
        P.~Vogel$^3$, Y-F.~Wang$^4$, J.~Wolf$^1$}
\address{$^1$ Department of Physics and Astronomy, University of Alabama, Tuscaloosa AL 35487 \\
         $^2$ Department of Physics and Astronomy, Arizona State University, Tempe, AZ 85287 \\
         $^3$ Division of Physics, Mathematics and Astronomy, Caltech, Pasadena CA 91125 \\
         $^4$ Physics Department, Stanford University, Stanford CA 94305}

\date{\today}
\maketitle
\begin{abstract}
The analysis and results are presented from 
the complete data set recorded at Palo Verde between September 1998 and July 2000.
In the experiment, the $\nuebar$ interaction rate  has been measured
at a distance of 750 and 890~m from the reactors
of the Palo Verde Nuclear Generating Station
for  a total of 350 days, including 108 days with one of the
three reactors off for refueling.
Backgrounds were determined by (a) the $swap$ technique based on 
the difference between signal and background
under reversal of the positron and neutron parts of the correlated event
and (b) making use of the conventional reactor-on and reactor-off cycles.
There is no evidence for neutrino oscillation and the mode 
$\nuebar\rightarrow\bar\nu_x$ was excluded at 90\% CL for
$\dm>1.1\times10^{-3}$~eV$^2$ at full mixing, and $\sinq>0.17$ at large $\dm$.
\end{abstract}
}

\section{Introduction}
We report here the
final results of a long baseline study of
$\nuebar$ oscillations at the Palo Verde Nuclear Generating Station. 
This is a continuation of the work
reported earlier in \cite{Boehm:1999gk,Boehm:1999gl} in which
details of the experiment and first results are described.
Hence we only briefly
describe  the detector and the analysis,
stressing the improvements and final results. 
Since the previous report, the data sample has been almost doubled.
Improvements have been made on reconstruction and simulation,
reducing the systematic error by one-third.

The experiment was originally motivated by the observation of an anomalous
atmospheric neutrino ratio $\nu_\mu/\nu_{\rm e}$
reported in several independent experiments
\cite{Fukuda:1994mc,Becker-Szendy:1992ym,Peterson:1999dc}.
The mass parameter suggested by this
anomaly is in the range of
$10^{-2}<\dm<10^{-3}$ eV$^2$
for two-flavor neutrino oscillation.
The Palo Verde experiment, together with the \textsc{Chooz} experiment
\cite{Apollonio:1999ae,Apollonio:1998xe} with a similar baseline,
were able to exclude  $\nu_\mu\rightarrow\nu_{\rm e}$
oscillations as the dominant mechanism for the atmospheric 
neutrino anomaly. 
While the experiment
has pursued its goal of exploring the then unknown region of
small $\dm$, recent data from Super-Kamiokande\cite{Fukuda:1998mi}
favor the $\nu_\mu\rightarrow\nu_\tau$ oscillation channel
over $\nu_\mu\rightarrow\nu_{\rm e}$.

\section{Experimental technique}

\subsection{Detector and signal}

The experiment was performed at
the Palo Verde Nuclear Generating Station in Arizona.
The plant consists of three identical
pressurized water reactors with a total thermal power of 11.63~GW.
The detector was located 
 at a shallow underground site, 890~m
from two of the reactors and
750~m from the third.
The 32~meter-water-equivalent overburden
entirely eliminated any hadronic component of cosmic radiation
and reduced the cosmic muon flux.

The segmented detector \cite{Boehm:1999gl}
consisted of 66 acrylic cells filled with 
11.34~tons of Gd-loaded liquid scintillator\cite{Piepke:1999db}.
A 0.8~m long oil buffer at both ends of each 9 m long cell
and a 1~m buffer filled with water (105~t) surrounding the
central detector
shielded it from
radioactivity originating in the photomultiplier tubes (PMTs)
and laboratory walls as well as from neutrons produced by cosmic
muons passing outside of the detector.
The outermost layer of the detector was an active muon
veto counter, providing 4$\pi$ coverage.
All materials
used in the construction of the detector and the laboratory were screened
for there radioactivity content by means of low background gamma ray
spectroscopy in order to control detector background.

The $\nuebar$ flux was detected via the correlated 
positron and neutron sub-events
from the reaction
$\nuebar$p$\rightarrow$ne$^+$.
The sub-events are
(1) the positron's kinetic energy ($<E> \simeq 2.4$~MeV) and two prompt annihilation
$\gamma$'s and
(2) the subsequent (delayed with a time constant
of $\sim 27\mu$s ) $\gamma$'s from
capture  of the thermalized neutron on Gd (with energy $\sim$ 8 MeV).

The data acquisition  (DAQ) electronics was built as a dual bank system, 
allowing both parts of the sequential $\nuebar$ capture event
to be recorded with no deadtime by switching between banks.
Signals from each PMT were discriminated by two thresholds: a
{\em high} threshold corresponding to
$\sim$600~keV for energy deposits in the
middle of the cell and a {\em low} threshold
corresponding to $\sim$40~keV also in the
middle of the cell, or a single photoelectron (SPE) at the PMT.
The trigger processor, a field programmable gate array,
searched for {\em triple} patterns in the
central detector for each of the sub-events, 
requiring one high and at least two low
discriminator signals from neighboring cells
\cite{Gratta:1997cy}. All events with two {\em triple}
signals within 450~$\mu$s 
of each other were written to disk.

A veto signal following the passage of a muon
(typical veto rates were $\sim$2~kHz)
disabled the central
detector trigger for 10 $\mu$s.
With each event, the time and hit pattern of the previous muon
in the veto counter was recorded for offline use
along with information as to whether or not the muon passed through the
target cells. 
The veto inefficiency was measured to be
2.5$\pm$0.2\% for stopping muons (one hit missed) and 0.07$\pm$0.02\%
for through-going muons (two hits missed).

In order to maintain constant data quality during running,
a protocol of continuous calibration and monitoring of all
central detector cells was followed.
Relative timing and position were calibrated
with blue light emitting diodes (LEDs) installed inside each cell.
Additional blue LEDs  illuminated
optical fibers at the end of each cell,
providing information about PMT linearity
and short term gain changes. LED and fiber optic scans
were performed once a week.
For absolute energy calibration and 
determination of the positron and neutron detection efficiencies, 
as well as mapping of the
light attenuation in each cell,
radioactive sources were used.
A complete source scan was undertaken every 2--3 months.
Further details have been described in Ref. \cite{Boehm:1999gl}.

\subsection{Expected $\nuebar$ interaction rate}

To evaluate the expected $\nuebar$ interaction rate
in the detector, the power and fuel composition of the
three reactors must be known. The calorimetric methods,
based on the measurement of temperature and water flow-rate
in the secondary cooling loop, provided power determination with
0.7\% uncertainty.

The fission rates in the three reactor cores were calculated 
daily using a simulation code provided by the
manufacturer of the reactors. 
The output of the core simulation had been checked
by measuring isotopic abundances
in expended fuel elements in the core;
errors in fuel exposure and isotopic abundances are estimated
to cause $<0.3$\% uncertainty in the $\nuebar$ flux estimate.
Four isotopes  ---
$^{239}$Pu, $^{241}$Pu, $^{235}$U, and $^{238}$U ---
produce virtually
all the thermal power as well as all the $\nuebar$'s.
Measurements of the neutrino yield per fission and energy spectra
exist for the first three isotopes\cite{Hahn:1989zr,Schreckenbach:1985ep}.
The $^{238}$U yield, which contributes 11\%
to the final $\nuebar$ rate, was calculated from theory\cite{Vogel:1980bk}
with an uncertainty of 10\%. The contribution of $^{238}$U  
to the uncertainty of the
total neutrino rate is therefore $\sim$1\%.

The $\nuebar$ energy spectrum  was reconstructed from the
measured positron  kinetic energy.
The approximate relation $E_{\nuebar} = E_{e^+}$ + 1.8 MeV is slightly
modified by the kinetic energy carried away by the neutron
($\sim$50~keV). The cross section of the detection reaction
is accurately known \cite{Vogel:1999zy}; the dominant uncertainty
(0.2\%) stems from the neutron lifetime.

Previous short baseline reactor experiments have found good
agreement between calculated and observed neutrino fluxes 
\cite{Zacek:1986cu,Achkar:1995,Declais:1995su}. In particular,
Ref. \cite{Declais:1995su} quoted an uncertainty in the neutrino
flux per fission of 1.4\%.
Together with the combined uncertainty of 1.5\% of the reactor power,
the distance to the detector and the number of target atoms, the total
systematic uncertainty of the $\bar{\nu}_e$ interaction rate 
therefore amounts to 2.1\%.

\begin{figure}[htb!!!]
\centerline{\epsfxsize=3.5in \epsfbox{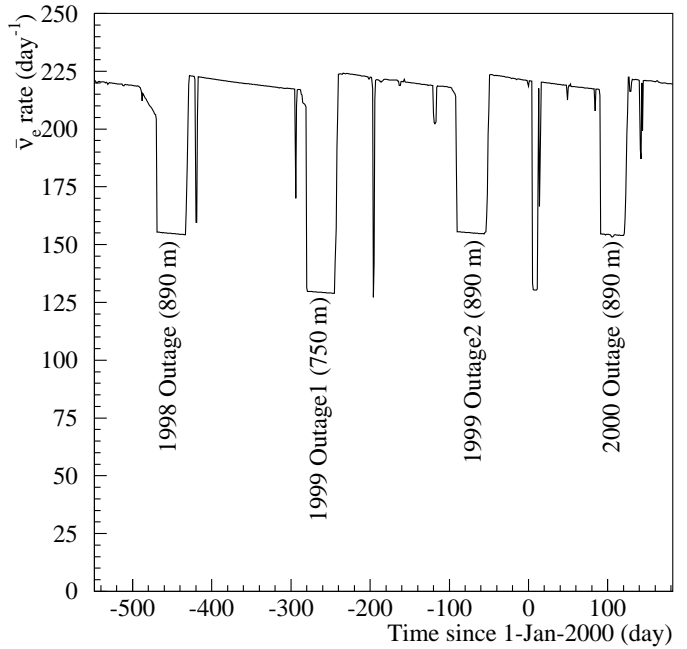}}
\caption{The calculated $\nuebar$ interaction rate in the detector
target for the case of no oscillations.  
The four long periods of reduced flux from reactor refuelings
were used for background subtraction.
The decreasing rate during the full power operation is a result
of the changing core composition as the reactor fuel is burned.}
\label{fig:rates}
\end{figure}

The expected $\nuebar$
interaction rate in the whole target,
both scintillator and the acrylic cells, is plotted
in Fig.~\ref{fig:rates} for the case of no oscillation from
July 1998 to July 2000.
Around 220 interactions per day are expected with all three units
at full power.
Four periods of sharply reduced rate occurred when one
of the three reactors was off for refueling,
the more distant reactors contributing each approximately
30\% of the rate and the closer reactor the
remaining 40\%. The short spikes of decreased rate are due to
accidental reactor outages, usually less than
a day. The gradual
decline in rate between refuelings is caused by fuel burn-up,
which changes the fuel composition in the core and the relative
fission rates of the isotopes, thereby affecting slightly the 
yield and spectral
shape of the emitted $\nuebar$ flux.

\section{Monte Carlo simulation}

\subsection{Detection efficiency}

Accurate understanding of the $\nuebar$ efficiency
is crucial. Therefore,
as described in \cite{Boehm:1999gl}, two parallel and independent
event reconstruction and detector simulation Monte Carlo
codes have been developed.
Both give consistent results; in  \cite{Boehm:1999gl}
the reported results were based on one of them, here most
of the results are based on the second method. 

A Monte Carlo model with a detailed simulation
of the detector response, including the PMT pulse shape,
is essential to simulate the rather strong 
dependence of 
the $\nuebar$ efficiency 
on the event location in the detector and, to a lesser extent,
on time due to some scintillator aging.
A variety of measurements was performed
to cross check the Monte Carlo  modeling of
the detector response.

The simulation code \cite{Boehm:1999gl}
contains the whole detector geometry and simulates
the energy, time, and position
of energy deposits in the detector using  \textsc{geant} 3.21\cite{geant}.
\textsc{gfluka}\cite{fluka} is used
to simulate hadronic interactions, while for the low energy neutron transport
\textsc{gcalor}\cite{gcalor} is employed.
Scintillator light quenching, parameterized as a function
of ionization density,
is included in the simulation\cite{birks}.

Given the output of the
physics generators, the
Monte Carlo simulates the
detector response in the form of PMT pulses
which are converted into time and amplitude digitizations
and trigger hits. Digitized data are then reconstructed with the
same programs as real data, providing the trigger and selection cuts efficiencies.

\begin{figure}[htb!!!]
\centerline{\epsfxsize=3.7in \epsfbox{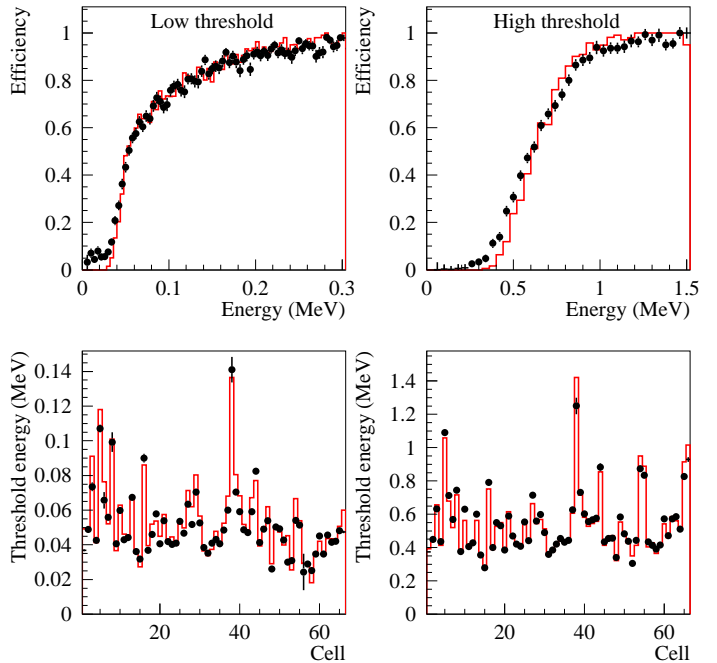}}
\caption{The upper plots show the
simulated and measured trigger efficiency for low and high thresholds as a
function of energy deposited in the center of one cell. Dots represent
data while the solid line shows the simulated efficiency.
The lower plots show the energy corresponding to a trigger efficiency of
50\% for each cell. The spread between data and Monte Carlo has been
improved by a factor of about two compared to \protect\cite{Boehm:1999gl}.}
\label{fig:threshold}
\end{figure}

\subsection{Improvements of the simulation}

Since the initial
results were published \cite{Boehm:1999gl},
the data sample has been almost doubled.
There were also improvements in the analysis
due to refinements of the simulation of the detector response.
Three changes had the largest impact on the quality of the simulation:
\begin{itemize}
\item The pulse-shapes of several hundred single-photoelectron (SPE) signals 
were digitized and compared with each other.   An average SPE pulse-shape was 
deduced, replacing the simple model that used only fixed rise- and decay-times.
\item 
The scintillation light was traced through the cell to the PMTs. 
SPE pulse-shapes with constant charge-to-amplitude ratio were added up 
to the final PMT pulse for each photo-electron produced in the cathode. 
The trigger threshold was compared to the amplitude of the total PMT pulse.
However, the charge-to-amplitude ratio
 of the measured SPE pulses varied slightly 
 from pulse to pulse, resulting in a smeared 
 trigger threshold when plotted as a function of ADC counts. Instead of varying
 the width of the average SPE pulse in the simulation,
 the relative height of the threshold was sampled
 from a Gaussian distribution with $\mu = 1$ and $\sigma$ adjusted to describe the
 measured slope of the trigger-efficiency versus ADC-counts.  The resulting modeling
 improvements can be recognized by comparing the trigger threshold accuracy in 
 Fig.~\ref{fig:threshold} with the analogous quantities in Fig.~9 of~\cite{Boehm:1999gl}).
\item The constant thresholds used for all cells and runs were replaced by an
 individual threshold for each discriminator. Variations in time were also taken
 into account by tracking thresholds
 using  neutrino runs.
\end{itemize}
In conclusion the new simulation reduced the spread of data/simulation over all
cells in the detector (Fig.~\ref{fig:threshold}) from 19.2\% to 10.2\%
for the low (SPE) threshold and from 7.6\% to 3.5\% for the high threshold.

\subsection{Test of the $\nuebar$ detection efficiency}

$^{22}$Na and Am-Be sources were used to verify
the absolute efficiency of the detector for positron annihilations and
subsequent neutron captures.
The 1.275 MeV primary $\gamma$ of the  $^{22}$Na source
is accompanied 90\% of the time by a
low energy positron which annihilates in the source capsule.
The primary $\gamma$  mimics
the positron ionization 
associated with a low energy $\nuebar$ event and, together
with the  annihilation $\gamma$'s, closely approximates
the positron portion of a $\nuebar$ event near the trigger threshold.

\begin{figure}[h!!!]
\centerline{\epsfxsize=3.7in \epsfbox{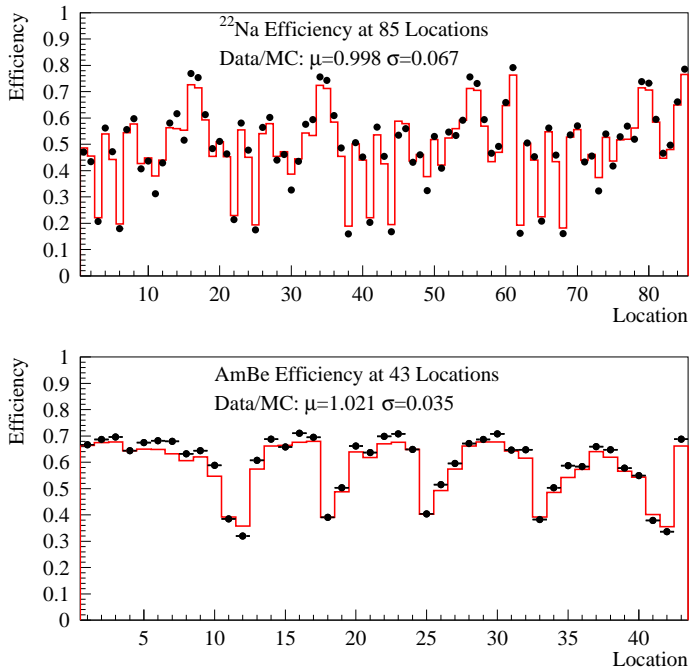}}
\caption{Comparison of data (points) and Monte Carlo (histograms) for
detection efficiency for $^{22}$Na and Am-Be source runs at various locations.
For positions of the radioactive source near the border of the central
detector we measure lower efficiencies in good agreement with the
simulation (see locations 3, 5, 22, ... for $^{22}$Na or locations
11, 12, 18 ... for Am-Be).}
\label{fig:eff}
\end{figure}

The $^{22}$Na source was
inserted into the central detector 
at various locations during four dedicated calibration
periods, separated by several months. A total of
85 different runs were taken
in order to sample various distances from the PMTs
and edges of the fiducial volume. 
This allows determination of an absolute efficiency
since the source activity is known to 1.5\%. 
After applying loose cuts to suppress background and
correcting for detector deadtime,
the measured absolute trigger efficiency could be
compared with the Monte Carlo prediction;
the results are shown in the top panel of Fig.~\ref{fig:eff}.
Good agreement is seen in the average efficiency over all runs
(the spread in data/Monte Carlo has been improved to 6.7\% in 85 locations
compared to 11.1\% in 36 locations in  \cite{Boehm:1999gl}),
and the agreement between the four calibration periods 
was better than 1.4\%.
The  $^{22}$Na energy spectra predicted by the simulation
and measured in the data  also agree well.
This comparison tests all aspects of the simulations:
the $high$ and $low$ trigger thresholds, and the
total energy deposit.

\begin{figure}[h!!!]
\centerline{\epsfxsize=3.7in \epsfbox{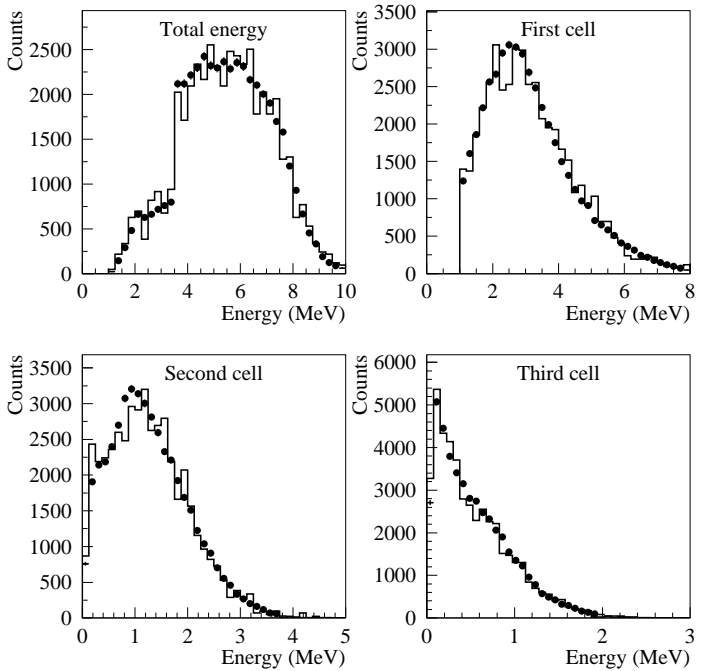}}
\caption{Comparison of data (points) and
Monte Carlo simulation (histograms) 
for the spectra of total energy and first, second, and third
most energetic hit ($E_{\rm total}$, $E_1$, $E_2$, and $E_3$)
for capture cosmic muon induced neutrons.
The sharp feature at 3.5 MeV total energy is related to the 
requirement that at least one sub-event has total energy above
this value (see text).}
\label{fig:esam}
\end{figure}

In order to check the neutron capture detection efficiency,
the Am-Be neutron source was attached to one end of a thin (7.5mm) 
NaI(Tl)-detector readout by a flat PMT, so that the entire assembly 
could be still inserted in the gaps between each cell and its 
neighbor above or below.  The NaI(Tl)-detector tagged the 4.4 MeV 
$\gamma$ emitted in coincidence with a neutron.

The NaI(Tl) tag forced the digitization of the 4.4~MeV $\gamma$ as the 
prompt part of an event and opened a 450~$\mu$s window for neutron 
capture, the same coincidence window as used in the $\nuebar$ runs.

Loose neutron cuts were applied, and
corrections were made for detector deadtime
and a low rate of random background.
On average, the Monte Carlo efficiency predictions agree
well over the 43 locations 
tested (compared to 25 locations in  \cite{Boehm:1999gl})
with an average agreement of better than 2.1\%, as shown in the
lower panel of Fig.~\ref{fig:eff}.
 Small adjustments of parameters
of the detector simulations could improve the agreement
of the AmBe efficiencies, but at the same time led to larger
disagreement for other parameters, e.g. $^{22}$Na efficiencies or
the shape of energy spectra. Therefore, only directly measured parameters
(trigger efficiency as a function of charge and shape of PMT pulses)
were used to adjust Monte Carlo parameters. Measured efficiencies for
$^{22}$Na and AmBe were only used to estimate the systematic uncertainty
of the simulation.

Again, the energy spectra
for neutrons predicted by the simulation
and measured in the data were compared. The total
energy seen in all cells and the energy detected in the
three most energetic hits is plotted in Fig.~\ref{fig:esam}.
This test was done with cosmic muon induced neutrons, which
are the dominant type of  correlated events in  neutrino runs.
The neutrons were equally distributed over the whole detector.

These procedures completely
test our $\nuebar$ efficiency simulation.
Thus, our ability to accurately generate the
events, model the detector response, reconstruct the events, and
correctly calculate the livetime of the
data acquisition (DAQ) system was verified.

The Monte Carlo simulation  yielded an average efficiency over the
whole detector as a function of $\nuebar$ energy.
The simulation included interactions in
the acrylic walls of the cells, since there is significant efficiency
for inverse beta decay originating there.
Next, the efficiency from the simulation was folded with the
incident $\nuebar$ spectrum (including possible distortions due to 
oscillations) to get the overall efficiency which generally depends 
on the oscillation parameters $\Delta m^2$ and $\sin^2 2\theta$.

\section{Analysis and Results}

We briefly discuss here event reconstruction, event selection,
efficiencies, and backgrounds.
Details may be found in \cite{Boehm:1999gk,Boehm:1999gl}.   
Both the analysis presented here and the one used for our previous
papers were repeated {\it without changing selection cuts} for the
present data-set.  

The energies and positions associated with hits were reconstructed for 
each bank.      The position of the hit
along the length of the cell was determined from TDC times with a 
time--walk correction applied on the basis of the collected charge.
The collected charge for each end was corrected for light attenuation
and PMT non--linearity and converted to an energy using energy
calibration constants.  The hit energy was determined as
the weighted average of the measurements from either end.

To select events in the energy ranges where
the triggers are efficient, we required that each sub-event (prompt 
``positron'' and
delayed ``neutron'') have at least one hit greater than 1 MeV and at least two
additional hits with energy greater than 30 keV.  Any event with hits
greater than 8 MeV in either sub-event was discarded.
The magnitude and pattern of energy deposits in the prompt sub-event
were required to resemble what was expected from the kinetic
energy of the positron and its annihilation. 
(The annihilation $\gamma$'s each had to have energy less than 600 keV,
and together less than 1.2 MeV. This is the only cut
which treats the two sub-events asymmetrically.)   The prompt 
and the delayed sub-events 
were required to be correlated in space and time.  To further 
suppress backgrounds, an event was accepted if it started at least 
150 $\mu$s after the
last veto hit and at least 3.5 MeV of energy was deposited in either 
the prompt or delayed sub-event. 

The event yield must then be corrected for the efficiency of trigger 
and selection cuts as well as for detector deadtime, which has
two components.  The first one is the loss of neutrino events
due to muons crossing the detector (a) within 150 $\mu$s before the start
of the neutrino event or (b) between the prompt and delayed sub-events. 
Its magnitude was determined from the measured muon veto rate
and the distribution of inter-event times from detector 
simulation.  The experiment livetime after losses due to the muon veto
is approximately 66\%.
The second deadtime component comes from the DAQ system being unavailable 
to digitize a triple.  Its magnitude
is the ratio of the number of triples for which the DAQ was busy
to the total number of triples seen by the trigger which could
be directly measured using scalers.
The deadtime of the trigger itself was measured to be less than 0.1\%.  
The experimental DAQ livetime was about 81\% for 1998 and 92\% for
1999--2000.   The higher DAQ livetime in 1999--2000 was due to recording 
correlated events only, rather than all triples, thus strongly reducing 
the load on the DAQ system.    For the case of no oscillations, the 
combined efficiency of the trigger and selection cuts on neutrino 
interactions is about 18\%.  The detector deadtime further reduces
the efficiency to about 10\% (the exact figure for each period being
given in Table~I).

Experimental backgrounds may be naturally classified as {\em{uncorrelated}}
and {\em{correlated}}, with  uncorrelated backgrounds  due to 
random coincidences between triple triggers within the delayed 
coincidence window, and correlated background due to
events in which both sub-events belong to the same process.   

The dominant source of {\it uncorrelated} events
is natural radioactivity. The inter-event time distribution 
for uncorrelated background 
events follows an exponential function with a time constant of $\sim$500
$\mu$s, as would be expected given the muon veto rate of $\sim$2 kHz
and the veto--dependent event selection requirements.  This time 
dependence is slow compared to that of signal and correlated 
backgrounds, hence the uncorrelated background could be separated and
studied by looking at long inter-event times. 

The main source of {\it correlated} background are neutrons from muon 
spallation or capture.  These events are mainly comprised of 
proton--neutron events in which a single neutron deposits
its kinetic energy by scattering from protons and is then 
captured, and double neutron events in which two (typically
thermal) neutrons from the same spallation event are captured in
the detector. 

\begin{table*}
\caption{Data taking periods, efficiencies (including livetime), 
measured event rates $N_1$ and $N_2$, and results of the {\em swap}
analysis (see text), including the various background estimates.
Uncertainties are statistical only.}
\label{tab:rates}
\begin{tabular}{lcccccccc}
Period  & \multicolumn{2}{c}{1998} & \multicolumn{2}{c}{1999-I} & 
 \multicolumn{2}{c}{1999-II} & \multicolumn{2}{c}{2000} \\
Reactor & on & 890 m off & on & 750 m off & on & 890 m off & on & 890 m off \\
\hline
time (days)      & 30.4 & 29.4 & 68.2 & 21.8 & 60.4 & 29.6 & 83.2 & 27.5 \\
efficiency (\%)  & 8.0  & 8.0 & 11.5 & 11.6 & 11.6 & 11.6 & 10.9 & 10.8 \\
\hline 
 & \multicolumn{8}{c}{measured rates} \\
\hline
$N _{1}$ (d$^{-1}$)  & 39.6 $\pm$ 1.1 & 34.8 $\pm$ 1.1 & 54.9 $\pm$ 0.9 & 45.1 $\pm$ 1.4 
 & 54.2 $\pm$ 0.9 & 49.4 $\pm$ 1.3 & 52.9 $\pm$ 0.8 & 43.1 $\pm$ 1.3   \\
$N _{2}$ (d$^{-1}$) & 25.1  $\pm$ 0.9 & 21.8 $\pm$ 0.9 & 33.4 $\pm$ 0.7 & 32.0 $\pm$ 1.2 
 & 32.5 $\pm$ 0.7 & 32.6 $\pm$ 1.0 & 30.2 $\pm$ 0.6 & 30.4 $\pm$ 1.1  \\
$(1 - \epsilon_1)B_{\rm{pn}}$ (d$^{-1}$) & 0.88 & 0.89 & 1.11 & 1.11 & 1.11 & 1.11 & 1.07 & 1.07  \\ 
\hline
 & \multicolumn{8}{c}{efficiency corrected rates} \\
\hline
Background (d$^{-1}$)   & 292 $\pm$ 11 & 255 $\pm$ 10 & 265 $\pm$ 6 & 266 $\pm$ 10 
 & 256 $\pm$ 6 & 265 $\pm$ 9 & 249 $\pm$ 5 & 272 $\pm$ 9 \\
$R_\nu$   (d$^{-1}$)        & 202 $\pm$ 19 & 182 $\pm$ 18 & 212 $\pm$ 10 & 124 $\pm$ 17 
 & 214 $\pm$ 11 & 161  $\pm$ 15 & 237 $\pm$ 10 & 129 $\pm$ 16 \\
$R_{\rm calc}$ (d$^{-1}$)     & 216 & 154 & 218 & 129 & 220 & 155 & 218 & 154 \\
\end{tabular}
\end{table*}

Our analysis is based on 350.5 days of data taking, of which 242.2 days
were at full power  and the remainder at partial power with
a reactor down for refueling.  
For the analysis we 
subdivided the data into 8 periods.  Four of the periods 
correspond to the four reactor refueling periods in which one 
reactor was off ({\em{off}} periods).  Each of the remaining four periods 
({\em{on}} periods) are constructed
from intervals of full power data 
bracketing each refueling
period.  Table \ref{tab:rates} shows the running time for each of the 8 periods
and the distance to the down reactor for each of the {\em{off}} 
periods.

The raw trigger rates for triples and correlated triples were
approximately 50 Hz and 1 Hz, respectively.  For 1999--2000,
the typical event rate after selection was $\sim$55 d$^{-1}$ with all
reactors at full power.  Under the assumption of no oscillations, the 
efficiency after the trigger, deadtime, and event selection
for detecting $\bar{\nu} _e$'s above inverse beta decay
threshold  was $\sim$11\%; precise estimates of the efficiency 
period--by--period are listed in Table \ref{tab:rates}.  
The observed event rate, $N_1$, may be compared to an expected signal rate of
$\sim$25 d$^{-1}$ for no oscillations, implying a signal-to-noise ratio of $\sim 0.8$. 
The uncorrelated background event rate after selection was $\sim$7 d$^{-1}$.

\subsection{Analysis with the ``reactor power'' method}

From Table \ref{tab:rates} it is evident that the event rate is
significantly lower during each refueling period.

\begin{figure}[htb!!!]
\centerline{\epsfxsize=3.7in \epsfbox{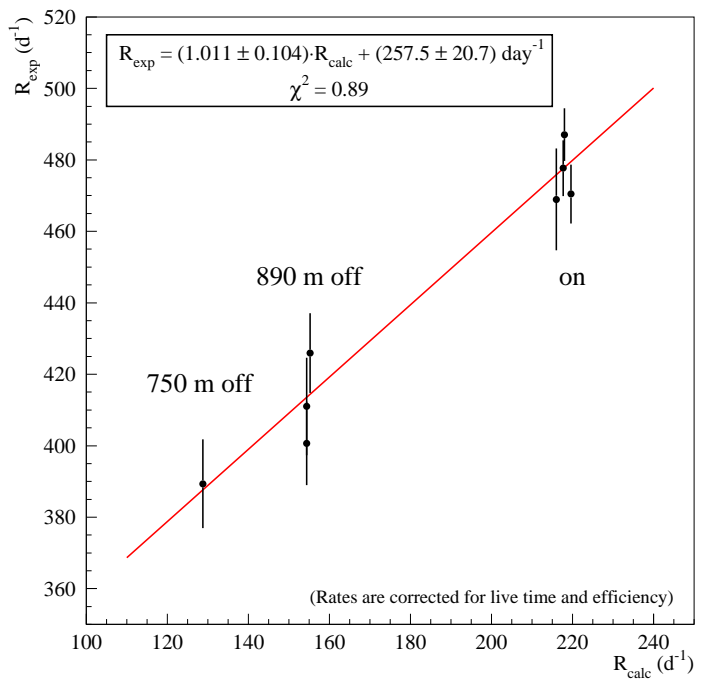}}
\caption{The event rates $R_{{\rm exp}}$ for different data taking periods, 
corrected for deadtime and neutrino
detection efficiency, plotted versus the expected neutrino interaction
rate $R_{{\rm calc}}$ for no oscillations.  Errors are statistical only. 
Points corresponding to data taking periods 
with same reactor power conditions should lie on top of each other.
Also shown
is the result, discussed in the text, of a linear fit to the data.}
\label{fig:ratefit}
\end{figure}

To investigate more quantitatively the correlation between event rates
and reactor power, we plot  in Fig. \ref{fig:ratefit} the experimental 
rate corrected for efficiency and deadtime $R_{{\rm exp}}$, against the 
calculated signal rate $R_{{\rm calc}}$ expected for no oscillations.
Only statistical uncertainties are indicated.  If the data were
consistent with no oscillations and the background were constant, then 
the points should lie along a straight line with unity slope.  The
y--intercept is equal to the rate of background interactions scaled by 
the ratio of the effective background detection efficiency to the
neutrino detection efficiency.
The data are in fact consistent with lying along a straight line.  A 
linear fit to these data gives a slope of 1.011 $\pm$ 0.104 (stat.)
and a y--intercept of 257.5 $\pm$ 20.7 (stat.) d$^{-1}$.  
The reduced $\chi ^{2}$ of the fit is 0.89.  Our data are therefore 
consistent with the hypothesis of no oscillations.

\begin{figure}[htb!!!]
\centerline{\epsfxsize=3.7in \epsfbox{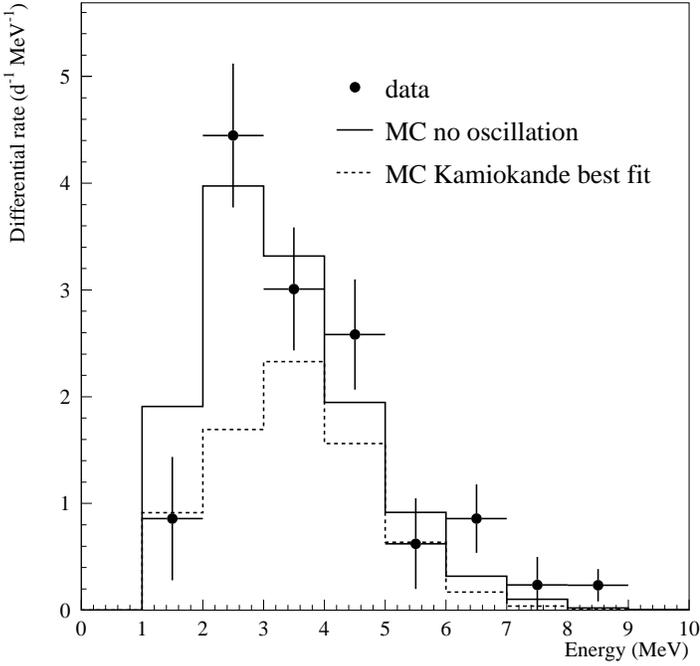}}
\caption{The prompt energy spectrum after {\em{on}}-{\em{off}}
subtraction averaged over the 4 pairs of 
{\em{on}}/{\em{off}} periods.
The histograms show the corresponding expectations for no oscillations
(solid line) and the Kamiokande best fit (dashed line).}
\label{fig:eonoff}
\end{figure}

We have also analyzed the energy dependence of the neutrino interactions
to see whether it is consistent with no oscillations as well.
For each of the four pairs of on--off periods, one may subtract the
event rate at partial power from the rate at full power.  The 
resulting difference, after the small correction for the
fuel burn-up has been made,  is the contribution to the full power event
rate from the neutrinos emitted by the reactor unit which was being
refueled during the off period.  Fig. \ref{fig:eonoff} shows the
measured on--off event rate difference binned in the visible 
prompt (positron) energy.  These data are not corrected for efficiency
or deadtime.
To make this plot, the weighted average
of the four pairs of on--off periods is taken.  Also shown are the
corresponding expectations from Monte Carlo for two scenarios: no
oscillations and oscillations with parameters obtained from the best
fit to the Kamiokande data \cite{Fukuda:1994mc}.  The comparison of our data
with Monte Carlo shows that the hypothesis of no oscillations is 
consistent not only with the measured event rate but also with the 
observed positron energy spectrum ($\chi^2/{\rm d.o.f}$ = 1.39
for 8 degrees of freedom), while it is not
consistent with the rates implied by the  Kamiokande best
fit parameters  ($\chi^2/{\rm d.o.f}$ = 3.69).

To test our data for oscillation hypotheses throughout the  
$\Delta m^2-\sin^22\theta$ plane for two flavor mixing, a $\chi ^{2}$ ana-\\
lysis using the ``reactor power'' changes was carried out.    
The $\chi ^{2}$ is defined as  
\begin{equation}
\chi^2=\sum_{i=1}^{8}
\frac{\left(R_{\rm exp}^{i} - bg - \alpha \cdot R_{\rm calc}^{i}\right)^2}{\sigma_{i}^2}+
\frac{(\alpha-1)^2}{\sigma_{\rm syst}^2}
\label{eq:chisqspec}
\end{equation}
where $R_{\rm exp}^{i}$ is the observed rate for period $i$, $bg$ is the background
rate, $R_{\rm calc}^{i}$ is the 
calculated rate for the period $i$ which depends on $\Delta m^{2}$ and $\sin ^{2} 2 
\theta$, and $\alpha$ accounts for possible global normalization
effects due to systematic uncertainties; $\sigma_{\rm syst}$ is 0.061
(see Table \ref{tab:syst} and the section on systematic errors below).
The quantity $bg$ is
scaled by 0.94 for the two periods in 1998 to account for the 
different trigger conditions in 1998 but is otherwise treated as
constant.  This scaling factor was determined from measuring how
the rates for double neutron background events (the selection of
which is described in the following section) and uncorrelated background
events changed between 1998 and 1999.

To define the 90\% confidence level (CL) acceptance region for
our data, we have followed the procedure suggested by Feldman and
Cousins (FC) \cite{Feldman:1998qc}.  We implemented this procedure in
two ways.  

The first, called {\em{Monte Carlo}} implementation, was realized as follows.  
A $\chi ^{2} _{\rm{best}}$ was determined by minimizing Eq. (\ref{eq:chisqspec}) 
with respect to $bg$, $\alpha$, $\Delta m ^{2}$, and $\sin ^{2} 2 \theta$
for physically allowed values of $\Delta m^2$ and $\sin ^{2} 2 \theta$. We 
found the best fit to correspond to a null $\sin ^{2} 2 \theta$ with a 
slightly un-physical $\alpha = 1.007$ and a $\chi ^{2} _{\rm{best}}/d.o.f.$ = 5.8/6 .  
The physical $\Delta m^2-\sin^22\theta$ plane was then subdivided into a fine grid.
At each grid point, we minimized Eq. (\ref{eq:chisqspec}) with respect 
to $bg$ and $\alpha$ to obtain 
$\Delta \chi ^{2} _{\rm{data}} \equiv
\chi ^{2} _{\rm{data}}(\Delta m ^{2},\sin ^{2} 2 \theta) - \chi ^{2} _{\rm{best}}$.  
For determining whether  the grid point was allowed at the 90\% CL,
we simulated 10$^{4}$ independent experiments at each grid point.  The same $\chi ^{2}$ 
minimization procedure was carried out for each simulated experiment
as for the data to obtain 10$^{4}$ $\Delta \chi ^{2} _{\rm{MC}}$'s.  
These $\Delta \chi ^{2} _{\rm{MC}}$'s were sorted in increasing order 
to find $\Delta \chi ^{2} _{\rm{c}}$, the value of $\Delta \chi ^{2}$ 
greater than 90\% of the $\Delta \chi ^{2} _{\rm{MC}}$'s.  If 
$\Delta \chi ^{2} _{\rm{data}} < \Delta \chi ^{2} _{\rm{c}}$, the 
grid point was accepted.

The second way in which we implemented the FC procedure, 
called {\em{raster scan}}, subdivides the two-dimensional grid in $\Delta m ^{2}$ slices.  
For each value of $\Delta m ^{2}$, Eq.~(\ref{eq:chisqspec}) is minimized with respect to 
$bg$, $\alpha$,  and $\sin ^{2} 2 \theta$.  The value obtained for
$\sin ^{2} 2 \theta$, without restricting the fit to the physically allowed range,
is denoted $(\sin ^{2} 2 \theta) _{\rm{best}}$ 
and its corresponding error is denoted $\sigma _{sin}$. 
The best fit is slightly unphysical; for $\Delta m ^{2} = 0.1$ eV$^2$,
$(\sin ^{2} 2 \theta) _{\rm{best}}/\sigma _{sin} = -0.2$.
The one-dimensional 90\% CL upper limit on $\sin ^{2} 2 \theta$ at the 
fixed value of $\Delta m ^{2}$ is then given by 
$\alpha _{\rm{FC}} 
\sigma _{sin}$ where $\alpha _{\rm{FC}}$ is looked up from Table X of
Reference \cite{Feldman:1998qc} for $x _{0} \equiv 
\frac{(\sin ^{2} 2 \theta) _{\rm{best}}}{\sigma _{sin}}$.  

While the raster scan method does not yield the global $\chi ^{2}$ 
minimum in the $\Delta m^2-\sin^2 2\theta$ plane, it is computationally 
much faster.  Checks have been carried out 
that the two methods for implementing the Feldman--Cousins procedure
yield the same limits.
For the purpose of determining
the regions of parameter space excluded by our data, knowledge of the 
$\chi ^{2}$ global minimum is not required.   We have therefore used 
the {\em{raster scan}} method to obtain the exclusion curves reported
in this paper.

The dashed curve in Fig.~\ref{fig:exclrastal} shows the 
region of $\Delta m^2-\sin^22\theta$ plane excluded at the 90\% CL
by our data analyzed with the ``reactor power'' method.  In the limit
of large $\Delta m^2$, the range $\sin^2 2\theta > 0.33$ is excluded;
whereas in the limit of maximal mixing, the range $\Delta m^2 > 
1.6 \times 10^{-3} ~\rm{eV} ^{2}$ is excluded.  We note that, in the
limit of large $\Delta m^2$, the Monte Carlo method excludes
the range $\sin^2 2\theta > 0.35$.

As already mentioned the independent analysis discussed in detail in 
\cite{Boehm:1999gk,Boehm:1999gl} was also improved and repeated for 
the full data-set.   In this case the ``reactor power'' analysis differs 
from the one described above in that the data are more finely binned by 
run rather than averaged by period.   There were typically two runs per day.  
After combining short runs (runs with fewer than 6 neutrino candidates) 
with adjacent runs, 698 data points were obtained.
A $\chi ^{2}$ analysis identical in approach to that described above 
was carried out using a systematic error that in this case amounts to 6.9\%. 
Again, the best fit is slightly unphysical; for $\Delta m ^{2} = 0.1$ eV$^2$
$(\sin ^{2} 2 \theta) _{\rm{best}}/\sigma _{sin} = -0.5$.   The 90\% CL 
exclusion contour obtained in this analysis is very similar to the dashed 
curve in Fig. \ref{fig:exclrastal}, but it is shifted toward smaller 
$\sin^2 2\theta$, with  $\sin^2 2\theta > 0.29$ excluded in the large 
$\Delta m^2$ limit.  The shift in the exclusion boundary 
is consistent with small systematic differences expected between the 
two independent reconstructions and analyses.

\subsection{Analysis with the ``swap'' method}

The ``swap'' method, where  the background is directly subtracted rather 
than using modulation of the reactor power, has substantially greater 
statistical power than the ``reactor power'' and, in addition, it has 
somewhat different systematics.

We briefly describe the method here; detailed descriptions have
already been published \cite{Boehm:1999gk,Boehm:1999gl,yfwang}.  
Let $N _{1}$ be the event rate after applying the neutrino selection cuts 
described above.   We then call $N_{2}$ the rate obtained by applying the
positron cuts to the delayed sub-events and the neutron cuts to the 
prompt sub-events (``swapped'' selection).
The measurements of $N_{1}$ and $N _{2}$ are listed for each
period in Table \ref{tab:rates}. It is found that only
about 20\% of the neutrino signal cancels in the difference $N_{1} - N_{2}$, 
as determined from Monte Carlo simulation. At the same time the uncorrelated
background and the double neutron component that dominates the correlated
background cancel in the difference.  We call $(1 - \epsilon_1)B_{\rm{pn}} =
(1 - \epsilon_{\rm sp})B_{\rm{pn,sp}} +(1 - \epsilon_{\rm cap})B_{\rm{pn,cap}}$ 
the residual contribution to $N_1 - N_{2}$, mainly due to the proton--neutron 
(``pn'')
component of the correlated background.  Here the $\epsilon$'s refer to 
the efficiency of the ``swapped'' selection for each channel~\cite{yfwang}, 
``sp'' denotes neutron production by $\mu$ spallation (mainly in the laboratory 
walls), ``cap'' by capture (mainly in the water buffer) and ``1'', maintaining the 
notation from our earlier papers, the total.  
While the latter process is well understood and can be reliably calculated with 
Monte Carlo simulations, the former is rather poorly known.  
The shape of the prompt energy spectrum for neutrons from spallation
was obtained generating neutrons in the laboratory walls according to several 
parametric models and passing them through the detector simulation and event 
selection.    The set of parametric models spanned the range of uncertainty in
our knowledge of the energy dependence of neutron production.
The normalization was then determined by assuming that 
high--energy neutrino--like events, selected by replacing the cut on maximum 
hit energy ($< 8$ MeV) in the ``positron'' sub-event by the requirement that 
at least one prompt hit have energy greater than 10 MeV, are due exclusively 
to spallation.   The uncertainty on the energy spectrum, quantified by the dispersion
between the different models, was taken into account in the systematic error.
The term $B _{\rm{pn,cap}}$ was found from the measured muon rate through the 
detector combined with the veto inefficiencies and the relatively well known 
total neutron production cross section and energy distribution by muon capture 
on oxygen. The main systematic derives from the veto inefficiency.

The quantity $(1 - \epsilon_1)B_{\rm{pn}}$ was estimated period by period, 
and the results are shown in Table \ref{tab:rates}.  The magnitude of 
$(1 - \epsilon_1)B_{\rm{pn}}$ is small compared to the difference 
$N _1 - N _{2}$, that is, the contribution from the proton--neutron component 
of the background largely cancels in the difference.  Therefore, 
even though the systematic error on $(1 - \epsilon_1)B_{\rm{pn}}$ is of order
100\%, the resulting contribution to the systematic error on the 
neutrino signal is only a few percent.
In the sixth line of Table~\ref{tab:rates} we list for each run-period
the resulting background (assuming for the purpose of this illustration
that the background efficiency is the same as for the signal, and correcting
for livetime).   The observed $\nuebar$ rate ($R_{\nu}$), corrected for 
livetime and efficiency, and the expected neutrino rate $R_{{\rm calc}}$ 
for no oscillations are also given in the Table.

Similarly to the reactor power analysis, 
we have carried out a $\chi ^{2}$ analysis to test our data for 
oscillation hypotheses throughout the two flavor oscillation
$\Delta m^2-\sin^22\theta$ plane.  The $\chi ^{2}$ definition is  
\begin{eqnarray}
\chi^2 & = &\sum_{i=1}^{8}
\frac{\left(N _{1,i} - N _{2,i} - (1 - \epsilon_1)B_{\rm{pn}} - 
\alpha (R_{\rm calc}^{1,i} - R_{\rm calc}^{2,i}) \right)^2}{\sigma_{i}^2}  \nonumber \\
& + &
\frac{(\alpha-1)^2}{\sigma_{\rm syst}^2}
\label{eq:chisqspec1}
\end{eqnarray}
where $\sigma_{\rm syst}$ for the ``swap'' method is estimated to be 0.053
as discussed below.    The free parameters in this definition of
the $\chi ^{2}$ are $\Delta m^2$, $\sin^22\theta$, and $\alpha$.
The Monte Carlo method gives $\chi ^{2}_{best}/d.o.f. = 10.3/7$ for 
$\sin^22\theta$ consistent with zero and, again, a slightly un-physical $\alpha = 1.008$.

\begin{figure}[htb!!!]
\centerline{\epsfxsize=3.7in \epsfbox{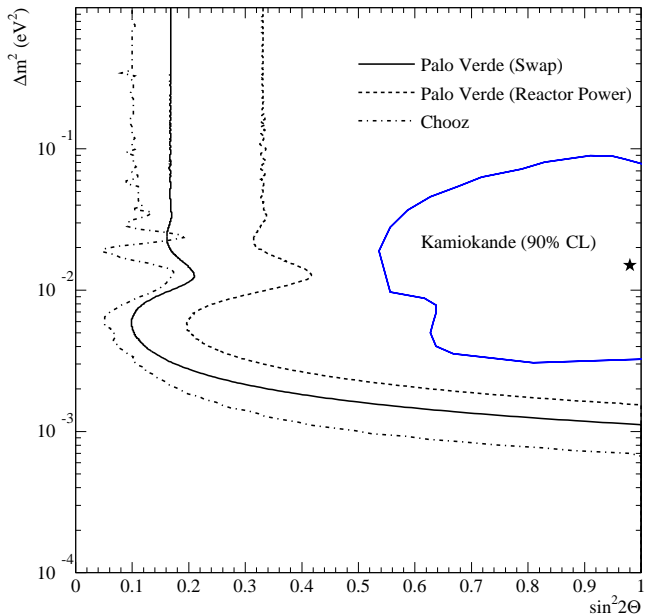}}
\caption{Regions of 
$\Delta m^2-\sin^22\theta$ plane (two flavor oscillations) excluded
at the 90\% CL by the ``reactor power'' analysis (dashed curve)
and ``swap'' analysis (solid curve).  Also shown are the Kamiokande
allowed region and best fit (star) and the region excluded by the Chooz
experiment \protect\cite{Apollonio:1999ae,Apollonio:1998xe}.}
\label{fig:exclrastal}
\end{figure}

The region of parameter space excluded at the 90\% CL by this analysis,
based on the raster scan method, is indicated by the solid curve in 
Fig.~\ref{fig:exclrastal}.    In the limit
of large $\Delta m^2$, the range $\sin^2 2\theta > 0.164$ is excluded;
whereas in the limit of large mixing, the range $\Delta m^2 > 
1.1 \times 10^{-3} ~\rm{eV} ^{2}$ is excluded.  We note that, in the
limit of large $\Delta m^2$, the Monte Carlo method excludes
the range $\sin^2 2\theta > 0.162$, and gives an essentially
identical exclusion curve.

\subsection{Test of the ``swap'' method}

As a further test of the ``swap'' method, we have investigated the energy 
dependence of $N _{1}$ and $N _{2}$.  The
measured energy dependence was compared to what would be expected
on the basis of our assumptions about the signal and background.  
Were a significant source of background ignored or incorrectly
treated, a discrepancy between data and expectation would 
result.

\begin{figure}[htb!!!]
\centerline{\epsfxsize=3.7in \epsfbox{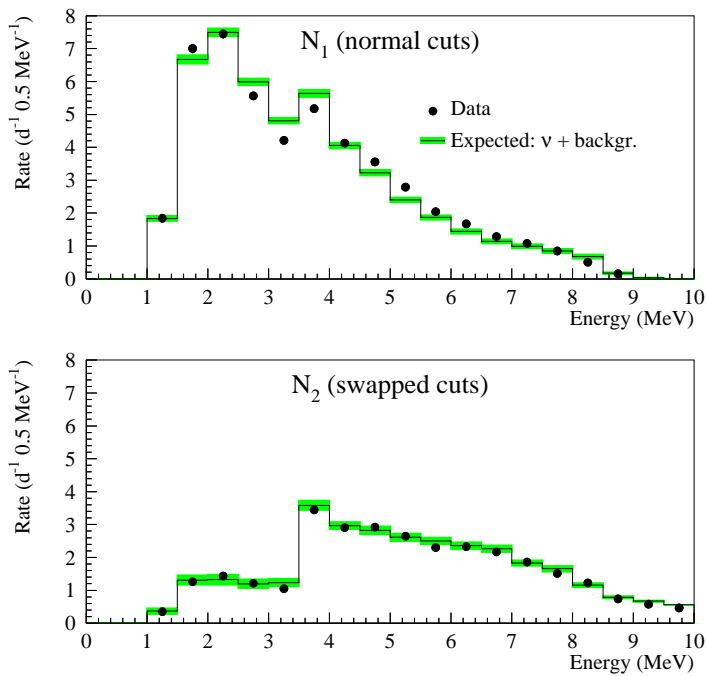}}
\caption{The energy spectrum of events comprising 
$N _{1}$ and $N _{2}$.  The points are measurements while 
the histogram shows the result of calculations described in 
the text.  The shading indicates the statistical uncertainty 
in the calculation, while error bars for the measurements are too 
small to be visible.}
\label{fig:eswap}
\end{figure}

To carry out this investigation, we assembled five samples of events:
\begin{itemize}
\item $\bar{\nu} _{\rm{e}}$: Inverse beta
decay events were generated and simulated in the detector with 
normalization determined from the reactor powers, cross section, and 
number of target protons.  No oscillations were assumed, as consistent
with the outcome of the {\em{reactor power}} analysis.
\item {\em{uncorrelated background}}: These events were selected from 
our data by
inverting the spatial and temporal correlation requirements between
the prompt and delayed sub-events. The data sample was normalized 
to reproduce the event rate at large inter-event times.  
\item $B _{\rm{pn,sp}}$: Neutrons produced by muon spallation in the 
laboratory walls were generated and passed through our detector 
simulation.  
As already mentioned, the sample was normalized 
 by assuming
that high--energy events satisfying the neutrino selection cuts 
are due to spallation.
\item $B _{\rm{pn,cap}}$: Neutrons produced by muon capture in  water
were generated and passed through the detector simulation.  This
data sample was normalized on the basis of the measured muon rate 
through the detector, the muon veto inefficiency,  
the fraction of muons stopping in the 
water, and the cross section for muon capture and neutron emission.
\item {\em{double neutron}}: Double neutron events were selected from data 
by requiring a muon hit within 100 $\mu$s preceding the start of the event
and applying the neutron capture cuts to both the prompt and 
delayed sub-events.  The sample was normalized so that--after 
application of the neutrino selection cuts--the combined 5 samples
gave the measured total $N _{1}$ rate.
\end{itemize}
The five data samples were subjected to the neutrino event selection
cuts ($N _{1}$) and the {\em{swap}} event selection cuts ($N _{2}$), 
respectively, and summed.  The resulting energy spectra, with statistical
uncertainties, are shown as histograms in Fig.~\ref{fig:eswap}.
The expectation from the sum of the five samples is in good agreement 
with the data (points).  Keeping in mind that only the overall normalization
of the $N_{1}$ spectrum is not independently determined, the validity of 
the ``swap'' analysis is solidly supported by this test.

\subsection{Systematic uncertainties}

The systematic uncertainty receives contributions from the detection
efficiency and the flux calculation.  In addition, the 
``reactor power method'' suffers a systematic error from background
variations with time, and the systematic uncertainty in the ``swap'' 
method has a contribution from the uncertainty in the estimate of 
$(1 - \epsilon_1)B_{\rm{pn}}$.

We have estimated the systematic
uncertainty in the detection efficiency as follows:
\begin{itemize}
\item {\em{$\nuebar$ selection cuts efficiency}}: The neutrino event selection 
cuts were varied randomly in the multi--dimensional cut space
over a reasonable range.  For each 
variation, the ratio of the observed number of events to that expected
for no oscillations was calculated. The systematic
error in the event selection efficiency was taken to be the {\em{rms}}
of the variations in the ratio. 
The uncertainties in the definition of the
energy scale are absorbed in this error component. 
The lower systematic uncertainty for the {\em{swap}}
method is due to cancellation of some systematics in the difference 
$N _{1} - N _{2}$.
\item {\em{$\rm e^{+}$ trigger efficiency}}:  The systematic uncertainty
in the $\rm e^{+}$ trigger efficiency is based on
comparison of simulated efficiencies with the measured efficiencies
for the $^{22}$Na calibration runs described above.  To decouple
uncertainties in 
the event selection efficiency from uncertainties in 
the trigger efficiency, loose
cuts designed to have negligible inefficiency were applied to select
$^{22}$Na events for this analysis.  The run--by--run 
comparison of the simulated and measured efficiencies has already
been shown in Fig. \ref{fig:eff}.  Averaged over all runs, the 
efficiencies agree to 0.2\% with a {\em{rms}} of 6.7\%.  Grouping the
runs by four calibration periods, the agreement was 1.3\%.  Combining 
this  with the 1.5\% uncertainty in the 
activity of the source, the estimated systematic error is 2.0\%.
\item {\em{n trigger efficiency}}: In an approach similar to that 
for estimating the systematic error in the 
$\rm e^{+}$ trigger efficiency, we have used the Am--Be calibration runs
described above to estimate the systematic uncertainty in the n
trigger efficiency.  The run--by--run comparison of measured versus
simulated n trigger efficiency has already been shown in Fig. \ref{fig:eff}.
The simulated efficiency is typically lower than the measured one.  This
difference is largely systematic, as manifested by the relatively small 
{\em{rms}} of 3.5\% across the different calibration periods and positions.
Averaged over runs, the difference between simulation and measurement is
2.1\%, which we assign as the systematic uncertainty for the n trigger
efficiency.
\end{itemize}

The results for these sources of systematic error for the two analysis
methods are listed in Table \ref{tab:syst}.
As explained in Section II.B the systematic uncertainty in the $\bar{\nu} _{\rm e}$ flux
is estimated to be 2.1\%.

The stability of background rates is a key assumption for the 
``reactor power'' analysis.  The actual level of background 
stability was estimated
by comparing the average rate (livetime--corrected) during the 
full power periods to the average rate during the partial power
periods for several background data samples: double neutrons, 
$B_{{\rm{pn}}}$, and uncorrelated background.  

In addition, Michel electron events, present in the data due to
the inefficiency of the veto detector, were used to track 
changes in the veto efficiency and, in particular,
in the background due to neutron production by muon capture in water.
These events are selected by requiring no activity in the muon detector, 
energy depositions in the prompt sub-event consistent with a muon track, and 
a delayed sub-event 5--20 $\mu$s later with an energy deposit of 10--70~MeV.  

The Michel electron data sample was observed to have a rate stability
better than 5\% and all other data samples were found to be stable 
to better than 1\%.  The rate variation for each background was 
normalized to
its estimated contribution to the neutrino event rate and then divided
by 12.9 $d^{-1}$, the average difference in livetime--corrected event
rates between full power and partial power periods.  
Combined, the resulting ratios for the four backgrounds indicated
a background instability of 2.1\% relative to the signal. We this take 2.1\%
as our estimate of the contribution
to the systematic error in the ``reactor power'' analysis from 
background variations.

The contribution to the uncertainty in 
$(1 - \epsilon_{\rm sp})B_{\rm{pn,sp}}$ from 
muon spallation in the walls was estimated from the spread in results from 
the four different models used to simulate neutron production and was 
found to be 0.29~$d^{-1}$.
The estimated contribution to the uncertainty in 
$(1 - \epsilon_{\rm cap})B_{\rm{pn,cap}}$
from the veto counter inefficiency resulted to be 0.94~$d^{-1}$.  Thus, 
the total systematic uncertainty on $(1 - \epsilon_1)B_{\rm{pn}}$
amounted to  0.98~$d^{-1}$.  This result was livetime-corrected and 
corresponded to 3.3\% of the average livetime--corrected value for
$N _{1} - N _{2}$.

The individual contributions are shown in Table~\ref{tab:syst} and added 
in quadrature to obtain the total systematic error for each analysis method.

\begin{table}
 \caption{Contributions to the systematic error of the ``reactor 
 power'' and ``swap'' analyses.} 
 \label{tab:syst}
  \begin{tabular}{lcc}
  Error source & ``reactor power'' (\%) & ``swap'' (\%) \\ \hline 
  e$^+$ trigger efficiency               & 2.0 & 2.0 \\ 
  n trigger efficiency                   & 2.1 & 2.1 \\ 
  $\bar{\nu}_{\rm e}$ flux prediction    & 2.1 & 2.1 \\ 
  $\bar{\nu}_{\rm e}$ selection cuts     & 4.5 & 2.1 \\  
  Background variation                   & 2.1 & N/A \\ 
  $(1 - \epsilon_1)B_{\rm{pn}}$ estimate & N/A & 3.3 \\ \hline
  {\bf Total}                            & {\bf 6.1} & {\bf 5.3} \\
  \end{tabular}
\end{table}

\section{Conclusion}

The results presented here, based on nearly double the number of events of
our previously published Palo Verde data, confirm the absence of
$\nuebar \rightarrow \bar{\nu}_x$ oscillations for low energy reactor neutrinos. 
The excluded regions for our ``reactor power'' and ``swap'' methods are 
enlarged accordingly.   For the ``reactor power'' method the new mixing 
angle limit is only slightly more restrictive than in our previous results.
This is due to a small shift of the central value of the fit.   
A substantially larger exclusion region is obtained with the ``swap''
method, thanks to reduced systematics.   In conclusion we find that the ratio
of observed interaction rate to the one expected for no oscillations is
$R_{\rm obs} / R_{\rm calc} = 1.01 \pm 0.024({\rm stat}) \pm 0.053({\rm syst})$.
These final results are dominated by systematics errors.

Our measurements, along with those reported by
Chooz~\cite{Apollonio:1999ae,Apollonio:1998xe}
and Super--Kamiokande \cite{Fukuda:1998mi}, excludes two
family ${\nu} _{\mu} - {\nu} _{e}$ mixing as being responsible for 
the atmospheric neutrino anomaly reported by Kamiokande \cite{Fukuda:1994mc}.

\section*{Acknowledgments}
We would like to thank the Arizona Public Service Company for
the generous hospitality provided at the Palo Verde plant.
The important contributions of M.~Chen, R.~Hertenberger, K.~Lou, and
N.~Mascarenhas in the early stages of this project are gratefully
acknowledged.
We are indebted to
J. Ball, B.~Barish, R.~Canny, M.~Dugger, A.~Godber,
J.~Hanson, D.~Michael, C.~Peck, C.~Roat, N.~Tolich,  A.~Vital,
and J.~Winterton
for their help.
We also acknowledge the generous financial help from
the University of Alabama, Arizona State University,
California Institute of Technology, and
Stanford University. Finally, our gratitude goes to CERN, DESY, FNAL,
LANL, LLNL, SLAC, and TJNAF who at different times provided us with
parts and equipment needed for the experiment.

This project was supported in part by the Department of Energy and
the National Science Foundation.
One of us (J.K.) received support from the Hungarian OTKA fund,
and another (L.M.) from the ARCS Foundation.


\end{document}